\documentclass[prb,twocolumn,showpacs,superscriptaddress]{revtex4-1}

\usepackage{amsmath, amssymb}    
\usepackage{graphicx}   
\usepackage{subfigure}
\usepackage{verbatim}   
\usepackage{color}      
\usepackage{subfigure}  
\usepackage{hyperref}   

\newcommand{\Eq}[1]{Eq.~(\ref{#1})}
\newcommand{\Fig}[1]{Fig.~\ref{#1}}

\newcommand{\cC}[0]{{\cal C}}

\begin{document}


\title{Geometric phases of d-wave vortices in a model of lattice fermions}

\author{Zhenyu Zhou 周振宇}
\affiliation{Department of Physics and Center for Materials Innovation, Washington University, St. Louis, MO 63130, USA}
\author{Oskar Vafek}
\affiliation{National High Magnetic Field Laboratory and Department of Physics, Florida State
University, Tallahassee, FL 32306, USA}
\author{Alexander Seidel}
\affiliation{Department of Physics and Center for Materials Innovation, Washington University, St. Louis, MO 63130, USA}


\date{\today}

\begin{abstract}
We study the local and topological features of Berry phases associated with the adiabatic transport of vortices in a d-wave superconductor of lattice fermions.
At half filling, where the local Berry curvature must vanish due to symmetries, the phase associated with the exchange of two vortices is found to vanish as well, implying that vortices behave as bosons. Away from half filling, and in the limit where the magnetic length is large compared to the lattice constant, the local Berry curvature gives rise to an intricate flux pattern within the large magnetic unit cell. This renders the Berry phase associated with an exchange of two vortices highly path dependent. However, it is shown that ``statistical" fluxes attached to the vortex positions are still absent. Despite the complicated profile of the Berry curvature away from half filling, we show that the average flux density associated with this curvature is tied to the average particle density. This is familiar from dual theories of bosonic systems, even though in the present case, the underlying particles are fermions.
\end{abstract}

\pacs{PACS}

\maketitle

{\em Introduction.}
A phenomenology based on a BCS-like pairing state with $d$-wave symmetry
has led to considerable success in understanding the properties of quasi-particles in 
high-T$_c$ superconducting cuprates. This includes the mixed state of these systems,\
where a magnetic field $H_{c_1}<H< H_{c_2}$ is applied and leads to the presence of
an Abrikosov vortex lattice.
Effective models\cite{WangPRB95, FranzPRL00, MarinelliPRB00, VafekPRB01a, VafekPRB01b, VafekPRL06, MelikyanPRB05, MelikyanPRB06, MelikyanPRB07,VafekPRL07, MelikyanPRB08, VishwanathPRL01, *VishwanathPRB02, TesanovicPRL04, Nikolic07, NikolicPRB06a}
have been developed that describe the dynamics of the quasi-particles under
 the simultaneous influence of magnetic field, the supercurrent flow due to the vortices,
and in some cases  the underlying microscopic lattice. 
The vortices of the mixed state are usually assumed to be static, i.e., frozen into
the Abrikosov lattice. However, it has been argued\cite{NikolicPRB06} that both
as a result of the small coherence length $\xi$, and possibly the proximity 
to an insulating state,  fluctuations of vortices may play a fundamental role.
Moreover,
at sufficiently high magnetic fields below $H_{c_2}$, it has been predicted 
that thermal and/or quantum fluctuations may melt the vortex lattice or glass, leading to a ``vortex liquid'' regime\cite{HuseNATURE92}.
For all these reasons, it is desirable to construct effective theories that include the vortices as
fundamental dynamical degrees of freedom. \cite{MizelPRB06}
Such a construction is readily available
in systems where the constituent particles are bosons, through the well-known Kramers-Wannier duality \cite{FisherPRB89}.
In Fermi systems, however, the vortex degrees of freedom only exist as dual partners of bosonic
Cooper pairs that are themselves emergent particles.
This arguably complicates the task of passing directly  from a microscopic description
in terms of electrons
to an effective
theory in terms of vortices, requiring more ad hoc assumptions.
Such effective theories have been previously discussed in a continuum formalism.\cite{NikolicPRB06}
In this paper, we aim to establish some key parameters of these  theories in
a microscopic lattice model. This is similar in spirit, but physically different, from earlier 
considerations for bosons in the absence of a lattice\cite{haldanewu}. 
These defining parameters include the quantum curvature felt by the vortices in the condensate,
that is, the effective magnetic field experienced by them, and their mutual statistics.
Specifically in the $d$-wave pairing case, where the continuum description of vortices
is somewhat plagued  by subtleties concerning self-adjoint extensions\cite{MelikyanPRB06},
our microscopic starting point also serves as lattice regularization\cite{VafekPRB01a},
which allows for a controlled study of the desired universal properties.

{\em Model description.}
We will study the 
Berry phases\cite{Berry1984} of vortices in the {\em BCS-Hofstadter} model, which has been used previously 
as a microscopic description of the mixed state in the cuprates\cite{VafekPRB01b}. 
\begin{equation}
\label{hamiltonian}
H = \sum_{\langle \bf rr' \rangle}[-t_{\bf rr'} c^\dagger_{\bf r\sigma} c_{\bf r'\sigma'}+\frac{\Delta_{\bf rr'}}{2} (c^\dagger_{\bf r\downarrow}c^\dagger_{\bf r'\uparrow}+c^\dagger_{\bf r'\downarrow}c^\dagger_{\bf r\uparrow})+h.c.]-\mu N
\end{equation}
In the above, 
the sum $\langle \bf rr' \rangle$ is over the nearest neighbors,
and  the hopping terms  are just those of the Hofstadter model, to be described below.
The corresponding uniform magnetic flux through the plaquettes of the lattice 
mirrors the fact that the penetration depth is much larger than the coherence length, as befits a type II
superconductor. Assuming symmetric gauge ${\bf A (r)}=(-y/2,x/2) \Phi$, where $\Phi$ the magnetic flux
per plaquette, the hopping amplitude assumes the form
\begin{align}
t_{\bf rr'} &=t\, e^{ -i {\bf \int_r^{r'} A }\cdot d{\bf l}}\,,
\end{align}
where ${\bf r }$ refers to the discrete sites of the lattice. The d-wave paring term is defined as 
\begin{align}
\Delta_{\bf rr'} & =\eta_{\bf r-r'} \Delta_{0, {\bf rr'}} e^{i \theta_{\bf rr'}}\label{drr} \\
\eta_{\bf r-r'} & =+/\!\!- \,\, \mbox{if} \,\, ({\bf r-r'})  \parallel \bf \hat x/ \hat y
\end{align}
Here, $\eta_{\bf r-r'}$ encodes the d-wave symmetry. $\Delta_{0, {\bf rr'}}$ is essentially constant, except for a suppression
of amplitude near the vortex core. We follow Ref. \cite{VafekPRL06}
in defining the pairing phase factor $e^{\theta_{\bf rr'}}$ via
\begin{equation}
\label{pairing phase}
e^{i\theta_{\bf rr'} }\equiv \frac{e^{i\phi(\bf r)}+e^{i\phi(\bf r')}}{\mid e^{i\phi({\bf r})}+e^{i\phi({\bf r'})} \mid}\,,
\end{equation}
i.e., as a link-centered average of a field $\phi(r)$ that satisfies the following continuum equations,
\begin{subequations}
\label{phiequations}
\begin{align}
& \nabla \times \nabla \phi ({\bf r})  =2\pi \hat z \sum_i \delta ({\bf r-r}_i) \\
& \nabla \cdot \nabla \phi(\bf r)  =0\,,
\end{align}
\end{subequations}
where the ${\bf r}_i$ denote the vortex positions, which can take on continuous values. The total number of vortices $n_V$ is equal to the number of half flux-quanta $\Phi_0$, $l_xl_y\Phi=n_V\Phi_0$ where $\Phi_0=\pi$ in natural units, and $l_x$, $l_y$ are the number of unit cells in the $x$ and $y$ direction, respectively.
Eq. \eqref{phiequations} can be solved\cite{MelikyanPRB07} via
\begin{equation}\label{theta}
\phi({\bf r})=\sum_i \{arg [ \sigma(z-z_i,\omega, \omega')]+2 \gamma (x-x_i)(y-y_i)+{\bf v}_0 \cdot {\bf r}\}\,,
\end{equation}
where  $\sigma(z,\omega, \omega')$ is the Weierstrass sigma-function with half periods $\omega=l_x/2$, $\omega'=il_y/2$, $z=x+iy$,
and the sum is over vortex positions. Integration constants ${\bf v}_0=2\sum_i {\bf A(r}_i)$ and 
$\gamma =\frac{\pi}{2l_xl_y}-\frac{\eta}{l_x}$
have been chosen such that the 
superfluid velocity  ${\bf v}_S=\nabla\phi/2-\bf A$ satisfies periodic boundary conditions, and averages to zero over the
magnetic unit cell\cite{VafekPRL06}, and $\eta=\zeta(\omega)$ is pure imaginary, with $\zeta$ the Weierstrass zeta-function.

The pairing phase factor $e^{i \theta_{\bf rr'}}$ in Eq.(\ref{pairing phase}) is ill-defined when the denominator goes to 0.
This is unacceptable since we mean to continuously change vortex coordinates  in the following.
To remove this singularity,
 we define $\Delta_{0,{\bf rr'}}$ as
\begin{equation}\label{core}
\Delta_{0,{\bf rr'}} \equiv \Delta_0 [1-\exp(-\frac{|e^{i\phi(\bf r)}+e^{i\phi(\bf r')}|}{\xi})]
\end{equation}
where $\Delta_0$ and $\xi$ are constant parameters. 
This leads to a suppression of pairing amplitude on links near the vortex, hence
$\xi$ may be thought of as a core radius.

We further impose periodic magnetic boundary conditions on our model as follows:
\begin{align}
\label{b.c.}
&c_{\bf r}  = T_x^{l_x}c_{\bf r}(T_x^{\dagger})^{l_x}=T_y^{l_y}c_{\bf r}(T_y^{\dagger})^{l_y} \\
&T_{\bf R}c_{\bf r}T_{\bf R}^{\dagger}=c_{\bf r+R}e^{i \int_{\bf r}^{\bf r+R} {\bf A} \cdot d{\bf l}+i {\bf R}\times {\bf r}\Phi}
\end{align}
In the above, the magnetic translation operators $T_x$ and $T_y$ are defined by letting ${\bf R}=\hat{x} \, or \, \hat{y}$.
We note that with the boundary conditions  \eqref{b.c.} imposed on electron operators, the physics
is also periodic in the vortex positions $\mathbf{r}_i$. That is, one may see that the formal
replacements $\mathbf{r}_i\rightarrow \mathbf{r}_i + l_x \hat x$, $\mathbf{r}_i\rightarrow \mathbf{r}_i + l_y \hat y$
affect the Hamiltonian by a unitary transformation, as given explicitly below. In particular, the quasi-particle spectrum
is invariant under such replacements.

{\em Calculation of the Berry phase.}
In the following, we will consider the model Eqs. \eqref{hamiltonian}-\eqref{b.c.}
as a function of vortex positions $\{\mathbf{r}_i\}$. We note that the simultaneous presence
of the magnetic field and the discrete ionic lattice generically opens up a gap in the
quasi-particle spectrum of the $d$-wave superconductor, except for special vortex
configurations that respect inversion symmetry \cite{VafekPRL07}.
The Berry phase associated with the motion of vortices is thus well-defined.
We further remark that the model defined above is traditionally studied
by means of a singular gauge transformation \cite{FranzPRL00}, 
that, on average, removes the magnetic field. This is inconvenient
for present purposes, since the precise transformation depends on 
vortex positions, and the Berry phase is clearly not invariant under 
unitary transformations that {vary} along the particular path in question. We thus need to
stay within the present framework of magnetic translations and associated boundary conditions.

To study the Berry phases associated with the motion of vortices, 
we first note that within our model the vortex positions 
are well-defined continuous parameters that are, at 
least for large enough lattice, entirely encoded in the pairing
amplitudes $\Delta_{\mathbf{rr}'}$. The Berry phase
associated with vortex motion along closed paths may be
computed via
\begin{equation}\label{berry}
e^{i\gamma}\approx \langle \Omega_1|\Omega_m\rangle \cdot \dots \cdot \langle \Omega_3 | \Omega_2\rangle \cdot \langle \Omega_2|\Omega_1\rangle\,,
\end{equation}
where the $|\Omega_i\rangle$ are the ground states of the system along a reasonably fine discretization 
of the path. The above formula has the advantage (over the standard integral formula) that 
a random, discontinuous phase that each $|\Omega_i\rangle$ acquires in numerical diagonalization 
automatically cancels.
Each ground state is constructed as the vacuum of Bogoliubov operators
\begin{align}\label{bogo}
&  \gamma_{n\uparrow}=\sum_{\bf r}(u_n^{*}({\bf r})c_{\bf r \uparrow}-v_n^{*}({\bf r})c_{\bf r \downarrow}^{\dagger} )\\
& \gamma_{n\downarrow}=\sum_{\bf r}(u_n^\ast ({\bf r}) c_{\bf r \downarrow}+v_n^\ast ({\bf r}) c_{\bf r \uparrow}^\dagger)
\end{align}
where the matrices $U_{{\bf r}n}=u_n(\bf r)$, $V_{{\bf r}n}=v_n(\bf r)$, satisfy Bogoliubov-deGennes equations
\begin{equation}
 \begin{pmatrix}
  -t-\mu & -\Delta\\
  -\Delta^* & t^*+\mu
 \end{pmatrix}
 \begin{pmatrix}
 U	\\
 -V
 \end{pmatrix}
 = E_n
 \begin{pmatrix}
  U	\\
 -V
 \end{pmatrix}
\end{equation}
for non-negative eigenvalues $E_n$.
It is clear from Eq. \eqref{bogo} that the state $ | \tilde{0} \rangle= \prod_{\bf r} c^{\dagger}_{{\bf r}\downarrow} |0\rangle$
is a vacuum of both the operators $\gamma_{n\uparrow}$ and $\gamma_{n\downarrow}^\dagger$,
where $|0\rangle$ is the vacuum of the $c_{{\bf r}\sigma}$ operators. The ground state of the Hamiltonian thus can be constructed as
\begin{equation}\label{Omega}
 |\Omega\rangle = \prod_n \gamma_{n\downarrow} | \tilde{0} \rangle\,.
\end{equation}
Using this last relation, and the inverse of Eq. \eqref{bogo}, one readily obtains 
\begin{equation}
\label{determinant}
\langle \Omega_i | \Omega_j \rangle  = \det (U_iU_j^{\dagger}+V_iV_j^{\dagger} )\,.
\end{equation}

{\em Results.}
We first consider the important special case of Eq. \eqref{hamiltonian} with $\mu=0$, or half-filling.
In this case the Hamiltonian is invariant under the {\em anti}-unitary charge conjugation 
operator defined via $\cC c_{{\bf r}\sigma} \cC =  (-1)^\mathbf{r} c_{{\bf r}\sigma}^\dagger$,
and the unique ground state $|\Omega\rangle$ is then invariant under $\cC$ as well
(up to a phase that can be made trivial). It then follows directly from \Eq{berry} that 
$e^{i\gamma}=\pm 1$. The first immediate conclusion from this is that 
as long as vortices are moved along contractible paths, the Berry phase must be
$+1$ for continuity reasons. If vortices were hard-core particles this would, in principle, still
leave the possibility of fermionic statistics. However, careful examination shows that the
Hamiltonian can be analytically continued without difficulty into configurations were two vortices
fuse into a double vortex at a given location. Exchange paths are thus contractible,
and hence vortices must satisfy bosonic statistics.
We have tested this for various lattice sizes and exchange paths. The model does, however,
become singular when vortex positions are formally approaching lattice sites, see \Eq{theta}.
It is thus possible that lattice sites carry an effective $\pi$-flux felt by vortices encircling such sites.
We have carefully checked that this is not the case in our model. Hence at half filling,
all Berry phases are unity. The above observations also hold for the $s$-wave case.

The observation that vortices are bosonic is non-trivial, since time reversal symmetry is absent,
and hence generically in two spatial dimensions even non-Abelian statistics are possible,
as is the case if the pairing symmetry is $p+ip$\cite{IvanovPRL01}.
Indeed, when we move away from half filling, there is no longer any symmetry that
requires the Berry phase to be trivial. We will now show that this situation
leads to a very intricate landscape of non-trivial quantum curvature.

\begin{figure}
\centering
\includegraphics[scale=0.38]{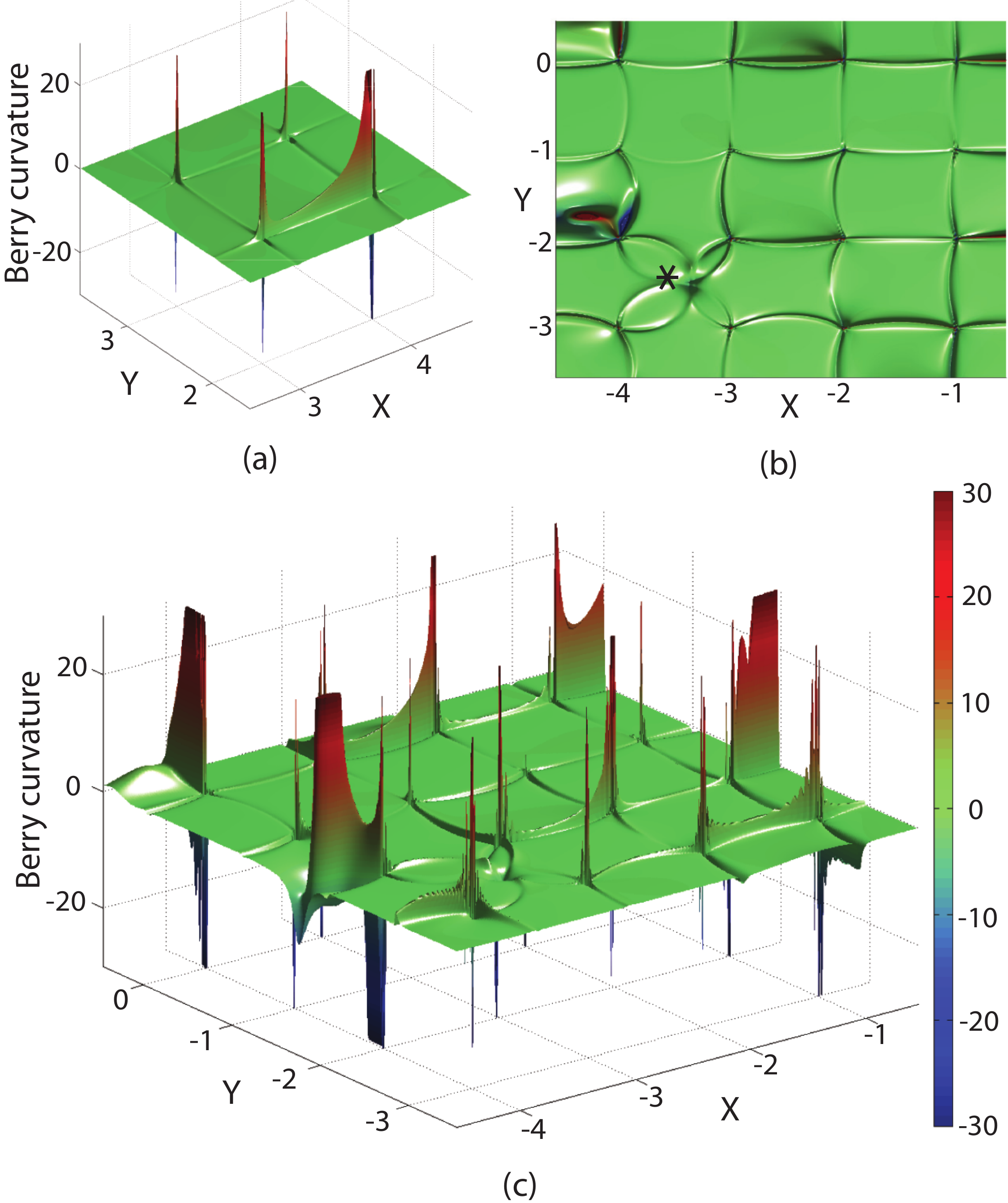}
\caption{(Color online) Berry curvature for 12 by 10 lattice in the presence of two vortices, for $\mu=0.05$. 
The calculated curvature at a given location is associated with the motion 
of one of the vortices at this location, while the position of other vortex is held fixed.
(a): 3D view of the Berry curvature in the vicinity of one plaquette.
Lattice sites and links are clearly visible, since the curvature is chiefly
concentrated on the latter, with singularities at the sites. The 
remaining space within the plaquettes is nearly flat.
(b) Top view of the lattice. The position of the fixed vortex is noted by an asterisk.
No singularity is present at this position. However, an interesting structure reminiscent of
the shape of a $d_{x^2-y^2}$ orbital appears in the vicinity. 
(c): 3D view of the entire lattice. Detailed features appear highly irregular,
except for the fact that the curvature is everywhere concentrated on the sites and the links of the lattice,
and on the structure seen around the position of the fixed vortex. }
\label{curl}
\end{figure}

The Berry curvature is defined as the Berry phase around an infinitesimal 
area, divided by the size of this area. In the following, we consider a lattice
containing only two vortices in the presence of periodic boundary 
condition. One vortex remains fixed, while for any point within the
unit cell, we calculate the Berry curvature associated with the motion
of the other vortex according to \Eq{berry}.
The Berry phase around arbitrary loops can be obtained as the 
integral of the Berry curvature over the enclosed area.
The result for a $12\times10$  lattice at $\mu=0.05$ is presented in \Fig{curl}(a).
It is apparent that the Berry curvature in this model is a highly non-trivial
function of position for any $\mu\neq 0$. One observes that the curvature is
conspicuously concentrated on the links and the sites of the lattice,
even though the vortex positions themselves are formally not tied to the discrete lattice.
Singular structures form in particular around the lattice sites.
These are described by $B(\mathbf{r})\sim a_i\delta(\mathbf{r}-\mathbf{r}_i)+f_i(\theta)/r$,
where $B(\mathbf{r})$ is the curvature, and $\theta$, $r$ refer to polar coordinates with the lattice
site $\mathbf{r}_i$ at the origin. The parameters $a_i$ and the functions $f_i$ depend 
sensitively on details such as the lattice size, $\mu$, the site index $i$, and the position of the other, fixed vortex.
Yet another interesting feature is the structure seen in the vicinity of the fixed vortex, which is somewhat reminiscent of the shape of a $d_{x^2-y^2}$ orbital. However, this structure does not seem to be reflective of by the pairing symmetry, but rather more the lattice symmetry, as similar calculations for the s-wave case show. We note that again no singularity 
indicative of  a flux tube
carried by the fixed vortex appears in \Fig{curl}(b) at the position of this vortex.
This implies that we should still think of these vortices as bosons, which move in an effective background magnetic field.

The complex nature of these features and the strong sensitivity on model parameters
are likely yet another facet of the fractal nature of the physics of the Hofstadter model.
To wit, in view of the fractal nature of the wave-vector dependence of spectral features of the Hofstadter model, 
it is reasonable to expect that
the response to a spatially inhomogeneous perturbation (coupling to many different wave vectors)
 is characterized by complicated and possibly chaotic spatial modulations.
The addition of a pairing order parameter with vortices clearly represents such a perturbation.
Here we are mostly interested in how to reconcile the complex features
seen at $\mu\ne 0$ with the trivial ones seen at $\mu=0$.
It is clear that our ability to precisely define the vortex position on scales below the lattice constant
is dependent on conventions, even though in the present case a natural convention is available, since our ground states are naturally parameterized by the vortex positions in the continuum field \eqref{phiequations} used to define the Hamiltonian.
We have, however,  tested the robustness of the qualitative features shown in  \Fig{curl}
by varying the precise form of the pairing order parameter \Eq{drr}.
In particular, we have varied the core parameter $\xi$ in \Eq{core}, and
tested various alternative forms for \Eq{pairing phase}. We also introduced variations in the
boundary conditions described above. In all cases we found that the qualitative
features of the Berry curvature remained unaltered.
Although we believe that the curvature landscape of \Fig{curl}(c) is interesting in its
own right, it is appropriate to make this landscape subject to some coarse graining procedure.
It is interesting to ask whether such coarse graining leads to a recovery of one of the 
basic facts suggested by conventional wisdom about vortex-boson duality, namely
that the curvature discussed above is directly tied to particle (here, Cooper-pair) density.
We will show in the following that this statement is recovered when the Berry curvature is averaged over the
magnetic unit cell (as opposed to the, typically much smaller, lattice unit cell). 
To this end, we again consider a lattice containing only two vortices within a single magnetic unit cell,
subject to the boundary conditions \eqref{b.c.}.

Let the coordinates of the ``moving'' vortex be $\mathbf{r}=(x,y)$.
As remarked initially, a formal shift of $x$ by $l_x$ changes the Hamiltonian by a gauge transformation. We have
$H\rightarrow U_x^\dagger H U_x$ with $U_x=e^{i\pi(1-y/l_y)\hat{N}/2}$, where $\hat N$ is the particle number operator.
Analogous relations hold for $y\rightarrow y+l_y$, with $U_y=e^{i\pi(1+x/l_x)\hat{N}/2}$.
We now calculate the Berry phase associated with a rectangular path of dimensions $l_x$, $l_y$
around the lattice. We may then choose a ground state phase convention $|\Omega(\mathbf{r})\rangle$ along the 
path satisfying 
\begin{align}
|\Omega(\mathbf{r}+l_y \hat y)\rangle=U_y^\dagger |\Omega(\mathbf{r})\rangle \nonumber\\
|\Omega(\mathbf{r}+l_x \hat x)\rangle=U_x^\dagger |\Omega(\mathbf{r})\rangle\label{bcvortex}
\end{align}
along the
horizontal and vertical path segments, respectively. 
The consistency of Eq. \eqref{bcvortex} with the continuity of $|\Omega(\mathbf{r})\rangle$ along the path
follows from the observation that for any ground state, $U_y^{\dagger}U_x^{\dagger}U_yU_x|\Omega\rangle= |\Omega\rangle$.
The latter holds because $U_y^{\dagger}U_x^{\dagger}U_yU_x=\exp(i\pi\hat N)$ and because the ground state
\Eq{Omega} always has even particle number parity. We determine the Berry phase for the rectangular path
as the integral over the Berry connection, $\langle\Omega(\mathbf{r})|\nabla|\Omega(\mathbf{r})\rangle$, where we observe that
\begin{equation}
\begin{split}
 \langle\Omega(\mathbf{r}+l_x \hat x)|\nabla|\Omega(\mathbf{r}+l_x \hat x)\rangle&=\langle\Omega(\mathbf{r})|U_x\nabla U_x^{\dagger} | \Omega(\mathbf{r})\rangle\\
&= \langle\Omega(\mathbf{r})|\nabla|\Omega(\mathbf{r})\rangle+\frac{i\pi}{2l_y}\langle \hat N\rangle  \hat y\,.
\end{split}
\end{equation}
It is clear that only the last term survives a cancellation between the vertical path segments, giving $i\pi\langle \hat N\rangle/2$.
The same contribution is obtained from the horizontal segment. We thus obtain $\gamma= \pi \langle \hat N\rangle$,
or $2\pi$ times the number of Cooper pairs in the system, in agreement with general expectations based on duality 
arguments applied to Cooper pairs\cite{NikolicPRB06}. It is worth noting that the quantity $\gamma$, when expressed 
as an integral of the Berry curvature over the entire lattice, is formally reminiscent of a Chern number. It is not
truly a Chern number, though, since the boundary conditions \eqref{bcvortex} do not quite allow one to make contact
with one-dimensional vector bundles over the torus. Indeed, $\gamma$ is not quantized, as $\langle \hat N\rangle$ may take on arbitrary  values in $[0,2l_x l_y]$.
We note that the derivation above is independent of the pairing symmetry.

{\em Conclusion.}
The present study establishes several aspects of Berry phases associated with vortex motion
in a microscopic model of superconducting lattice fermions.
It is shown that these vortices behave as bosons which, away from half filling, are subject
to a non-trivial effective magnetic field. In an average sense, it has been shown that this 
effective field is tied to the density of Cooper pairs. This is expected based on boson-vortex duality,
and was seen to emerge here in a microscopic model of fermions. We emphasize that 
the simple relation between Cooper-pair density and effective field is only seen to emerge 
after averaging over a magnetic unit cell. This may be used to justify a direct proportionality
between Cooper pair density and Berry curvature in the long wavelength effective theory.
However, our results also indicate
 that care must be used in order to justify such a relationship in general.
On the one hand, this is true because of the relatively large non-uniformity
of the observed Berry curvature within the magnetic unit cell.
Moreover, in the presence of particle hole symmetry we have found that
the Berry phase associated with closed paths is always zero, and thus corresponds
to $\pi$ times the average enclosed particle number only for such paths that happen to enclose
an even number of lattice sites. In this case, the background field appearing in the
effective theory should clearly be zero, and should not follow the total Cooper pair density.
This result will be robust to small perturbations respecting particle hole symmetry,
and is thus true for a wide class of microscopic models.
We conjecture that the complex landscape of the Berry curvature away from half filling 
is a facet of the fractal properties of the Hofstadter model, and believe that it is worthy
of further investigation.

{\em Acknowledgments.}
This work was supported  by the National Science Foundation under Grant No. DMR-0907793 (ZZ and AS)
and by NSF CAREER award under Grant No. DMR-0955561 (OV).

\bibliography{bibvortex}
\end{document}